\documentclass[]{revtex4-1}
\usepackage{amsmath}
\usepackage[pdftex]{graphicx}
\usepackage{graphicx}
\usepackage{amsmath}
\usepackage{amssymb}
\usepackage{url}
\usepackage{bm}							

\begin{document}


\title{Gallium nitride L3 photonic crystal cavities with an average quality factor of 16,900 in the near infrared} 



\author{Noelia Vico Trivi\~{n}o}
\affiliation{Institute of Condensed Matter Physics, Ecole Polytechnique F\'{e}d\'{e}rale de Lausanne (EPFL), CH-1015 Lausanne, Switzerland}

\author{Momchil Minkov}
\email[]{momchil.minkov@epfl.ch}
\affiliation{Laboratory of Theoretical Physics of Nanosystems, Ecole Polytechnique F\'{e}d\'{e}rale de Lausanne (EPFL), CH-1015 Lausanne, Switzerland}

\author{Giulia Urbinati}

\author{Matteo Galli}
\affiliation{Dipartimento di Fisica, Universit\`{a} di Pavia, via Bassi 6, 27100 Pavia, Italy}

\author{Jean-Fran\c{c}ois Carlin}
\affiliation{Institute of Condensed Matter Physics, Ecole Polytechnique F\'{e}d\'{e}rale de Lausanne (EPFL), CH-1015 Lausanne, Switzerland}

\author{Rapha\"el Butt\'e}
\affiliation{Institute of Condensed Matter Physics, Ecole Polytechnique F\'{e}d\'{e}rale de Lausanne (EPFL), CH-1015 Lausanne, Switzerland}

\author{Vincenzo Savona}
\affiliation{Laboratory of Theoretical Physics of Nanosystems, Ecole Polytechnique F\'{e}d\'{e}rale de Lausanne (EPFL), CH-1015 Lausanne, Switzerland}

\author{Nicolas Grandjean}
\affiliation{Institute of Condensed Matter Physics, Ecole Polytechnique F\'{e}d\'{e}rale de Lausanne (EPFL), CH-1015 Lausanne, Switzerland}

\date{\today}

\begin{abstract}
\textit{Photonic crystal point-defect cavities were fabricated in a GaN free-standing photonic crystal slab. The cavities are based on the popular L3 design, which was optimized using an automated process based on a genetic algorithm, in order to maximize the quality factor. Optical characterization of several individual cavity replicas resulted in an average unloaded quality factor $Q=16,900$ at the resonant wavelength $\lambda \sim 1.3~\mathrm{\mu m}$, with a maximal measured $Q$ value of 22,500. The statistics of both the quality factor and the resonant wavelength are well explained by first-principles simulations including fabrication disorder and background optical absorption.} 
\end{abstract}


\maketitle 


Over the past few years, there has been a strong effort, both in academia and industry, to combine the well-established Si technology with the optoelectronic properties of direct bandgap compound semiconductors.\cite{bookdadgar,roelkens2010} The aim is to unify different optoelectronic devices, with various functionalities and operating wavelengths, on the same chip while offering a reduced footprint.
In this regard, GaN, AlN, InN and their ternary alloys are excellent candidate materials for such platforms thanks to their wide direct bandgap ranging from the UV to the infrared (IR) spectral range. Their unique optoelectronic properties are supported by the significant increase in the solid-state lighting market which mostly relies on III-nitride light-emitting diodes. Besides the design flexibility enabled by such a bandgap tunability, III-nitrides (III-N) possess a large second-order nonlinear susceptibility, which is highly desirable for second harmonic generation.\cite{lundquist1994,vecchi2004,xiongpernice2011} For instance, this should facilitate the integration of fluorescence-based biosensors working in the green spectral range together with devices operating in the near-IR. Beyond enhanced light-matter interaction phenomena, this material family offers additional features such as chemical inertness, high thermal stability, large mechanical resistance, making them well-suited for optomechanics,\cite{xiongpernice2012} and biocompatibility.\cite{stutzmann2006}

The development of high quality ($Q$) factor III-N based photonic crystal (PhC) cavities -- both in the two-dimensional (2D) PhC slab and one-dimensional nanobeam geometries -- was hindered by technological issues mainly arising from their mechanical hardness and the lack of an appropriate lattice-matched substrate. However, in recent years several groups have overcome such challenges by reporting the fabrication of III-N PhC cavities exhibiting comparatively large experimental $Q$ factors. At short wavelengths, values up to 6,300 have been demonstrated both for nanobeam cavities \cite{sergent2012} and for defect cavities in PhC slabs, \cite{vicotrivino2012,samgiao2012,arita2012} whereas at $\sim$1.55 $\mu$m,  $Q$ values up to 34,000  where recently shown for GaN-on-Si PhC cavities based on the width-modulated waveguide structure\cite{VicoTrivino2013,Roland2014} and up to 146,000 for AlN nanobeam structures deposited by sputtering.\cite{pernice2012}

A number of considerations should be made in view of applications. First, it is important to assess the fluctuations in the resonant wavelength and the $Q$-factor which are expected in the fabrication of nominally identical structures due to imperfections.\cite{Portalupi2011,Minkov2013} In that sense, for reproducible integration in photonic devices, the average quality factor is an important figure of merit, while fluctuations in both quality factor and resonant wavelength should be minimized. In addition, while both the 2D slab and nanobeam geometries hold great promise as platforms for investigating strong light-matter coupling at the nano-scale, in a longer term perspective, the need for scalability and integrability hints at the 2D slab structure as the most suitable choice, where cavities and waveguides can be fabricated on a single slab and arranged in a circuit-like fashion. To this purpose, the minimization of the spatial footprint of PhC cavities is also a crucial requirement. In previous works, either L7 or width-modulated waveguide cavities were adopted at both visible \cite{vicotrivino2012,samgiao2012,arita2012} and IR wavelengths \cite{Roland2014}. Due to the larger mode volume, these designs are systematically characterized by a $Q$-factor larger than their smaller siblings, but only at the cost of a sub-optimal footprint. Previously, we demonstrated a measured $Q$ of 2,200 in a smaller L3 GaN-on-Si cavity.\cite{VicoTrivino2013} 

Here, we pursue these objectives by taking advantage of the recent progress in the automated design of PhC defect cavities based on the genetic optimization of the $Q$-factor.\cite{Minkov2014} We optimize a L3 cavity, designed to operate at $\sim$1.3 $\mu$m, to a theoretical quality factor $Q=166,000$. We fabricate several replicas of this design and characterize them optically. We demonstrate high reproducibility and an average measured (unloaded) $Q$-factor of 16,900, with individual samples reaching $Q=22,500$. We quantitatively explain the measured data -- and in particular the gap to the expected theoretical $Q$-factor -- by deploying a model that involves the simulation of hundreds of disorder realizations of the optimal design, and includes material absorption.

\begin{figure}
\includegraphics[width = 0.45\textwidth, trim = 0in 0in 0in 0in, clip = true]{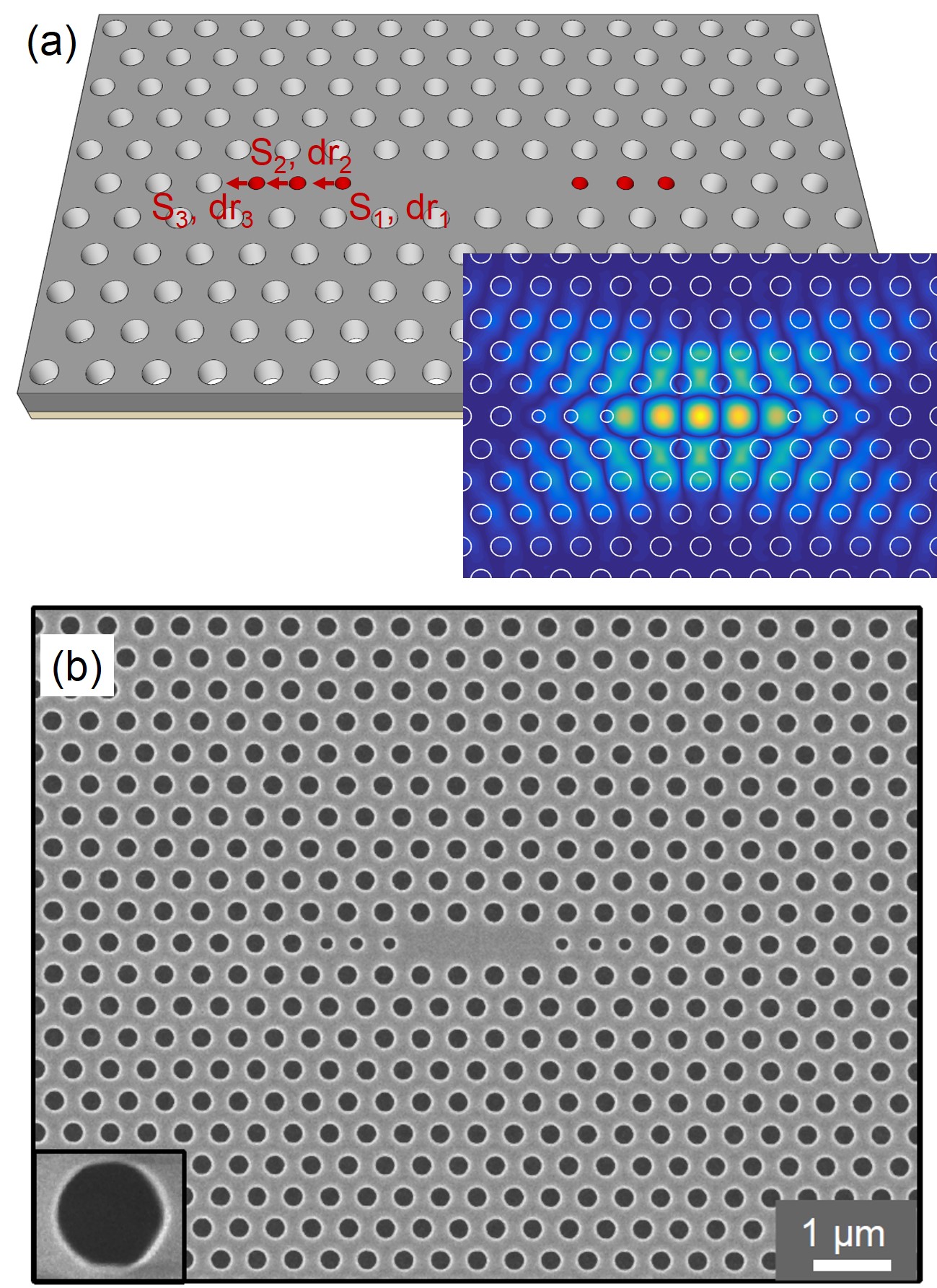}%
\caption{\label{fig1} (a): Schematic of the photonic crystal cavity design, with the AlN buffer layer shown in beige. The three holes which were modified for the optimized $Q$ are marked in red. Inset: simulated electric field ($|E_y|$) profile of the fundamental cavity mode. (b): SEM top view of a fabricated cavity. Bottom left: close-up view of one of the holes illustrating a trend to a hexagonal shape.}
\end{figure}

The PhC cavity design (Fig. \ref{fig1}(a)) that is optimized for high quality factor is that of the L3 cavity, consisting of three missing holes in a triangular lattice of cylindrical holes in a dielectric slab. The slab consists of 310 nm GaN (refractive index $n = 2.35$) and 40 nm AlN (refractive index $n = 2.05$), with the values of $n$ at 1.3 $\mathrm{\mu}$m extrapolated using Sellmeier formula coefficients given by Antoine-Vincent and co-workers for GaN on Si (111).\cite{vincent2003} The lattice constant is $a$ = 467 nm, while the hole radius was kept as a free parameter. The optimization was performed as outlined in detail in Ref. \onlinecite{Minkov2014}, using the guided-mode expansion (GME) method\cite{Andreani2006} to compute the $Q$ for a particular design and the genetic algorithm of the Matlab Global Optimization Toolbox \cite{matlab} to find the global maximum with respect to a given number of variational parameters. The method has already proven extremely effective in the case of silicon PhC cavities.\cite{Lai2014, Dharanipathy2014} In the case of silicon cavities, variations in the size of the holes were found to bring only a marginal improvement as compared to variations in their positions. The present problem however is different, particularly due to the much lower refractive index contrast, and we found that including size variations results in significantly better designs. Thus, both the positions and the radii of the three holes nearest to the cavity in the $\Gamma \mathrm{K}$ direction (Fig. \ref{fig1}(a)) were chosen as the parameters for optimization. The optimal design was found for the following parameters:  hole radius $r=0.2553a$, outward shifts of the three nearest holes $S_{1-3} = [0.3482, 0.2476, 0.0573]a$, shrinkage of their corresponding radii $dr_{1-3} = [-0.0980, -0.0882, -0.0927]a$. The high $Q$ of this optimized design was verified using a commercial-grade simulator based on the finite-difference time-domain (FDTD) method,\cite{lumerical} resulting in $Q = 112,000$ at a resonance wavelength $\lambda =$ 1.329 $\mathrm{\mu}$m. The simulated mode profile is plotted in the inset of Fig. \ref{fig1}(a). The value of the quality factor is more than 65 times larger than that of the corresponding L3 cavity with the same overall hole radius but with no variations ($Q = 1,700$). It is worth mentioning that all the simulations shown hereafter were performed with the FDTD solver. 

This high-$Q$ cavity design was then fabricated for experimental characterization. The PhC fabrication is first based on the growth on 2-inch Si (111) substrate by metal organic vapor phase epitaxy of a 40 nm thick AlN buffer layer followed by a 310 nm thick GaN layer. The PhC lattice is subsequently patterned by e-beam lithography using a SiO$_{2}$ hard mask and dry etching techniques. Finally, a membrane is obtained by undercutting the Si substrate, again by selective dry etching. With this processing methodology air gaps larger than $\approx 3 ~\mathrm{\mu m}$ can be achieved. Further details on this fabrication procedure together with structural characterization can be found in Ref. \onlinecite{VicoTrivino2013}. Figure \ref{fig1}(b) displays a scanning electron microscope (SEM) image of one of the fabricated L3 cavities. As shown in the bottom left inset, one can observe a slight deviation toward a hexagonal shape for the holes instead of the designed circular one. This is ascribed to a crystallographic orientation-dependent selective etching which has been previously reported in similar III-nitride based PhC lattice structures.\cite{Arita2007} In the present case, the cavity orientation is $[1\bar{2}10]$. Due to expected uncertainty in the etching, 20 groups of cavities (labelled $g_1-g_{20}$) were fabricated to allow for lithographic tuning. Group $g_{16}$ was targeted to have the nominal hole radius, while groups with lower (higher) number contain cavities where all radii are increasingly smaller (larger) in steps of 1 nm. 

\begin{figure}
\includegraphics[width = 0.45\textwidth, trim = 0.5in 0.2in 0.5in 0.3in, clip = true]{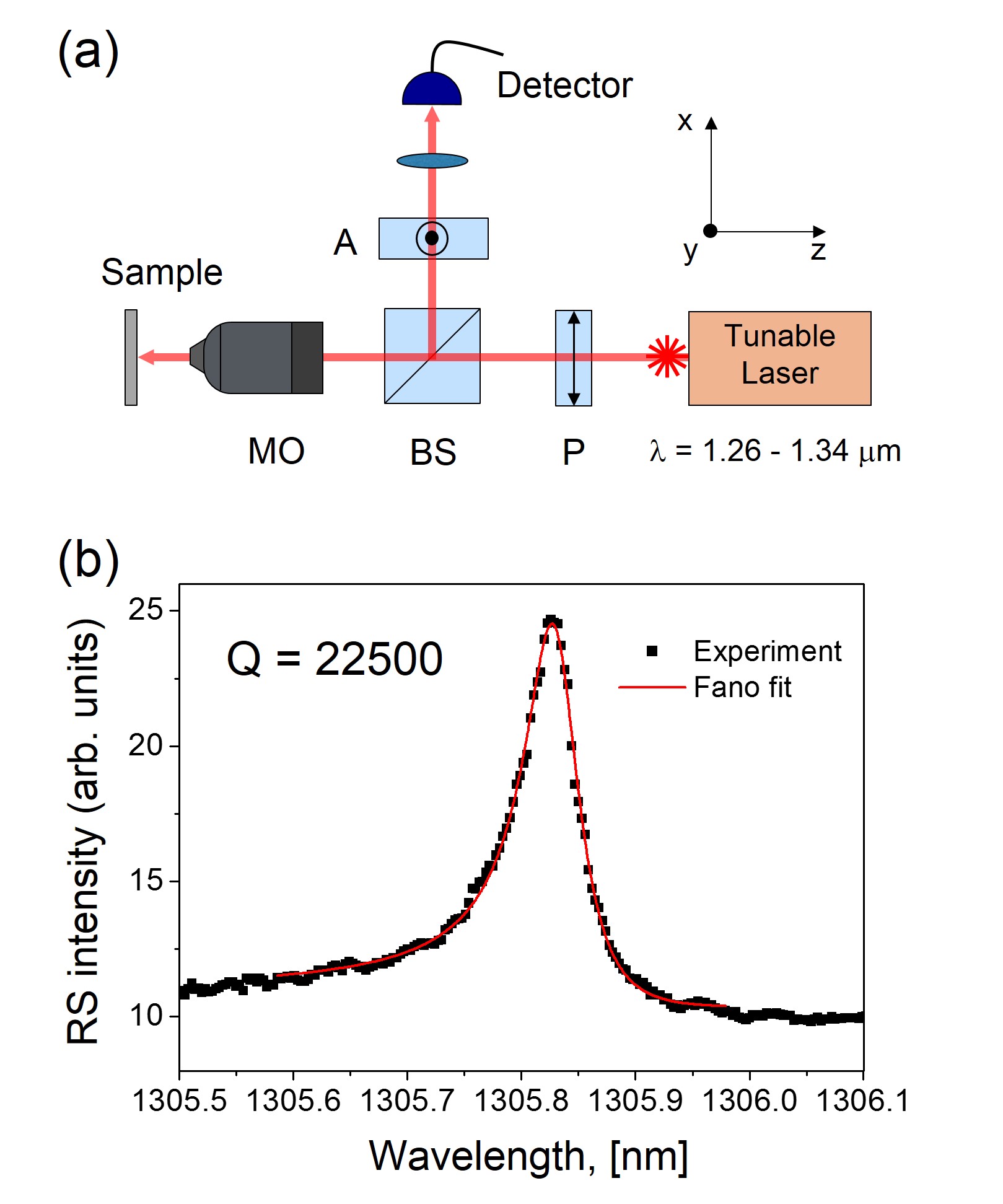}%
\caption{\label{fig2} (a): Schematic of the cross-polarization spectroscope; P: polarizer, BS: beam-splitter, MO: microscope objective, A: analyzer. (b): Resonant scattering spectrum and Fano fit of the cavity with the highest measured quality factor.}
\end{figure}

The spectroscopic characterization of the fabricated PhC cavities was carried out using cross-polarized laser light scattering. This technique has proven to be very effective in the experimental characterization of slab PhC cavities,\cite{Lai2014} as it yields the intrinsic (unloaded) quality factor and the precise resonance wavelength. The scheme of the experimental setup is shown in Fig. \ref{fig2}(a). Light from a continuous-wave tunable laser is linearly polarized by the polarizer P and focused on the sample by means of a high numerical aperture (NA = 0.8) polarization-maintaining microscope objective. The light reflected off the sample is collected using a beam splitter and analyzed by the analyzer A in cross-polarization with respect to the polarizer P. To maximize the signal from light which is resonantly coupled to the cavity mode, the cavity is positioned with its optical axis at 45$^\circ$ with respect to both the polarizer and the analyzer. Figure \ref{fig2}(b) shows a typical resonant scattering (RS) spectrum of a fabricated device with the highest measured $Q$-factor of $22,500$, where the characteristic asymmetric resonance is very well fitted by a Fano lineshape.    

\begin{figure}
\includegraphics[width = 0.45\textwidth, trim = 1in 3.5in 2in 5in, clip = true]{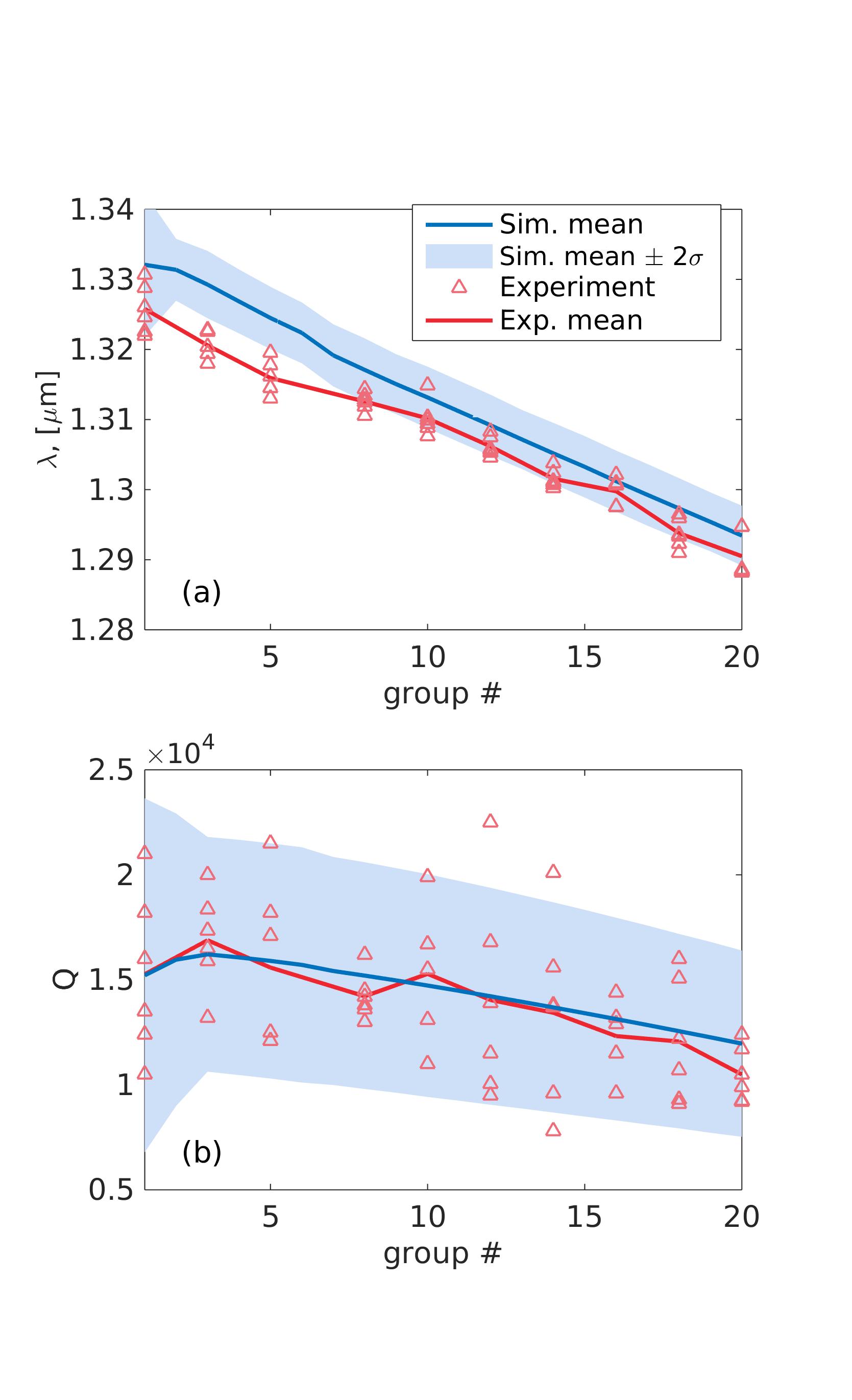}%
\caption{\label{fig3} Simulated (blue) and measured (red) (a) resonance wavelengths and (b) quality factors as a function of the cavity group number.}
\end{figure}

The measured resonance wavelength of the $g_{16}$ group of cavities, which was the target nominal structure, was blue-shifted by more than 20 nm from the simulated wavelength, 1.329 $\mathrm{\mu}$m. We attribute this shift to several possible effects: the deviation of the hole radius from the target value, the uncertainty in the slab thickness, and the uncertainty in the refractive index which comes from our Sellmeier's law extrapolation and potentially from free-carrier absorption. To fit the experimental data, the following assumptions were made: GaN refractive index $n = 2.33$, GaN layer thickness $d =$ 300 nm (the AlN layer was kept unchanged). In addition, a diameter increase of the holes occurring during the etching process was observed, which resulted in group $g_{10}$ having the nominal hole radius instead of $g_{16}$. To include the effect of disorder, we introduced a model with fluctuations in the hole positions and radii drawn from a uniform distribution with zero mean and standard deviation $\sigma_d =$ 5 nm. The model does not include the deviation of the holes from a circular shape (Fig. \ref{fig1}(b), inset), but is nonetheless a commonly used effective method to capture the statistics of disorder-induced losses.\cite{Portalupi2011, Minkov2012, Minkov2013} Using this model, 100 disorder realizations were simulated for each group $g_x$. In Fig. \ref{fig3}(a) we compare the experimentally measured and the simulated wavelengths $\lambda_{sim}$. The blue shaded area shows the region $\langle \lambda_{sim} \rangle \pm 2\sigma(\lambda_{sim})$, where the averaging is done over the 100 disorder realizations and $\sigma()$ denotes the standard deviation. The agreement is excellent given all the above-mentioned uncertainties. We note that the refractive index dispersion was neglected here as the GaN PhC structures are designed for a frequency range far below that of the GaN bandgap. A fine-tuning of the simulated results to the experimental ones is in principle possible, but would not bring any further insight. 
It is noteworthy that over the full set of groups the slope corresponding to the average of the experimental wavelength (red line in Fig. \ref{fig3}(a)) follows the expected theoretical trend (blue line) even when considering that there is a target hole diameter difference of only 2 nm between two adjacent groups ($g_x$ and $g_{x+1}$). This further confirms the reproducibility and maturity of this approach for the fabrication of III-nitride PhCs. 

The measured quality factors could also be well matched to our disorder simulation (Fig. \ref{fig3}(b)). To this purpose, constant, systematic losses with an associated quality factor $Q_a$ were additionally assumed. These could be due to absorption, but could also be related to scattering losses not captured in our disorder model, e.g. by the slab surface roughness. The value of $Q_a = 40,000$ was estimated for the best agreement between computed and experimental data. The simulated $Q_{sim}$ was computed as $Q_{sim}^{-1} = Q_d^{-1} + Q_a^{-1}$, where $Q_d$ is the quality factor resulting from the simulations in the presence of disorder, and the mean and standard deviation were obtained by $\langle Q_{sim} \rangle = 1/(\langle Q_d^{-1}\rangle + Q_a^{-1})$, $\sigma(Q_{sim}) = \langle Q_{sim} \rangle^2 \sigma(Q_d^{-1})$. The light-blue region in Fig. \ref{fig3}(b) is given by $\langle Q_{sim} \rangle \pm 2 \sigma(Q_{sim})$, and is thus the region within which, for a Gaussian statistical distribution, $95\%$ of the data points are expected to lie. The scattering in the measured data matches this distribution very satisfactorily. Finally, it is interesting to note that the quality factor slightly increases with decreasing radius, reaching a maximum around group $g_3$, where the theoretical $Q$ of the design with no disorder and no $Q_a$ is $166,000$. This is, however, followed by a drop and spread of the $Q$-distribution for smaller radii, for which, in some disorder realizations, a degradation of the cavity mode is observed (the field is no longer concentrated in the three-missing-hole region). 

In conclusion, we demonstrated a PhC L3 cavity optimized for fabrication in GaN, with a theoretical quality factor of $166,000$ at wavelengths in the $\sim$ 1.3 $\mathrm{\mu}$m window. Structures based on such a design were fabricated, incorporating lithographic tuning in the overall hole radius, and very good agreement was observed between the experimental data and our theoretical model in presence of disorder. Most importantly, the fabricated cavities consistently showed measured quality factors above $10,000$, and the maximum measured value was $Q = 22,500$. In addition, for the best group ($g_3$), the average measured $Q$ was $16,900$, which demonstrates the growing technological maturity of GaN-based PhCs. An additional advantage of the L3 design is that it is a point-defect cavity, thus it has a very small mode volume, which is suitable for strong light-matter interactions, and a very small footprint allowing for dense integration in two-dimensional circuits. Combined with the high quality factors demonstrated here, our design is of great promise for future applications.

\vspace{0.5cm}
\begin{acknowledgments}
This work was supported by the Swiss National Science Foundation through Projects N\textsuperscript{\underline{o}} 200020\_132407 and 200020\_150202.
\end{acknowledgments}


\begin{thebibliography}{10}%
\makeatletter
\providecommand \@ifxundefined [1]{%
 \ifx #1\undefined \expandafter \@firstoftwo
 \else \expandafter \@secondoftwo
\fi
}%
\providecommand \@ifnum [1]{%
 \ifnum #1\expandafter \@firstoftwo
 \else \expandafter \@secondoftwo
\fi
}%
\providecommand \enquote [1]{``#1''}%
\providecommand \bibnamefont  [1]{#1}%
\providecommand \bibfnamefont [1]{#1}%
\providecommand \citenamefont [1]{#1}%
\providecommand\href[0]{\@sanitize\@href}%
\providecommand\@href[1]{\endgroup\@@startlink{#1}\endgroup\@@href}%
\providecommand\@@href[1]{#1\@@endlink}%
\providecommand \@sanitize [0]{\begingroup\catcode`\&12\catcode`\#12\relax}%
\@ifxundefined \pdfoutput {\@firstoftwo}{%
 \@ifnum{\z@=\pdfoutput}{\@firstoftwo}{\@secondoftwo}%
}{%
 \providecommand\@@startlink[1]{\leavevmode}%
 \providecommand\@@endlink[0]{}%
}{%
 \providecommand\@@startlink[1]{%
  \leavevmode
  \pdfstartlink
   attr{/Border[0 0 1 ]/H/I/C[0 1 1]}%
   user{/Subtype/Link/A<</Type/Action/S/URI/URI(#1)>>}%
  \relax
 }%
 \providecommand\@@endlink[0]{\pdfendlink}%
}%
\providecommand \url  [0]{\begingroup\@sanitize \@url }%
\providecommand \@url [1]{\endgroup\@href {#1}{\urlprefix}}%
\providecommand \urlprefix [0]{URL }%
\providecommand \Eprint[0]{\href }%
\@ifxundefined \urlstyle {%
  \providecommand \doi [1]{doi:\discretionary{}{}{}#1}%
}{%
  \providecommand \doi [0]{doi:\discretionary{}{}{}\begingroup
  \urlstyle{rm}\Url }%
}%
\providecommand \doibase [0]{http://dx.doi.org/}%
\providecommand \Doi[1]{\href{\doibase#1}}%
\providecommand \selectlanguage [0]{\@gobble}%
\providecommand \bibinfo [0]{\@secondoftwo}%
\providecommand \bibfield [0]{\@secondoftwo}%
\providecommand \translation [1]{[#1]}%
\providecommand \BibitemOpen[0]{}%
\providecommand \bibitemStop [0]{}%
\providecommand \bibitemNoStop [0]{.\EOS\space}%
\providecommand \EOS [0]{\spacefactor3000\relax}%
\providecommand \BibitemShut [1]{\csname bibitem#1\endcsname}%
\bibitem{bookdadgar}%
  \BibitemOpen
  \bibfield{author}{%
  \bibinfo {author} {\bibfnamefont{T.}~\bibnamefont{Li}}, \bibinfo {author}
  {\bibfnamefont{M.}~\bibnamefont{Mastro}},\ and\ \bibinfo {author}
  {\bibfnamefont{A.}~\bibnamefont{Dadgar}},\ }%
  \emph{\bibinfo {title} {{III-V} Compound Semiconductors: Integration with
  Silicon-Based Microelectronics}}\ (\bibinfo {publisher} {{CRC} Press},\
  \bibinfo {year} {2010})\ ISBN \bibinfo {isbn}
  {9781439815229}\BibitemShut{NoStop}%
\bibitem{roelkens2010}%
  \BibitemOpen
  \bibfield{author}{%
  \bibinfo {author} {\bibfnamefont{G.}~\bibnamefont{Roelkens}}, \bibinfo
  {author} {\bibfnamefont{L.}~\bibnamefont{Liu}}, \bibinfo {author}
  {\bibfnamefont{D.}~\bibnamefont{Liang}}, \bibinfo {author}
  {\bibfnamefont{R.}~\bibnamefont{Jones}}, \bibinfo {author}
  {\bibfnamefont{A.}~\bibnamefont{Fang}}, \bibinfo {author}
  {\bibfnamefont{B.}~\bibnamefont{Koch}},\ and\ \bibinfo {author}
  {\bibfnamefont{J.}~\bibnamefont{Bowers}},\ }%
  \bibfield{journal}{%
  \Doi{10.1002/lpor.200900033}{\bibinfo {journal} {Laser Photon. Rev.}}\ }%
  \textbf{\bibinfo {volume} {4}},\ \bibinfo {pages} {751} (\bibinfo {month}
  {Nov.}\ \bibinfo {year} {2010}),\ ISSN \bibinfo {issn}
  {1863-8899}\BibitemShut{NoStop}%
\bibitem{lundquist1994}%
  \BibitemOpen
  \bibfield{author}{%
  \bibinfo {author} {\bibfnamefont{P.~M.}\ \bibnamefont{Lundquist}}, \bibinfo
  {author} {\bibfnamefont{W.~P.}\ \bibnamefont{Lin}}, \bibinfo {author}
  {\bibfnamefont{Z.~Y.}\ \bibnamefont{Xu}}, \bibinfo {author}
  {\bibfnamefont{G.~K.}\ \bibnamefont{Wong}}, \bibinfo {author}
  {\bibfnamefont{E.~D.}\ \bibnamefont{Rippert}}, \bibinfo {author}
  {\bibfnamefont{J.~A.}\ \bibnamefont{Helfrich}},\ and\ \bibinfo {author}
  {\bibfnamefont{J.~B.}\ \bibnamefont{Ketterson}},\ }%
  \bibfield{journal}{%
  \Doi{10.1063/1.112134}{\bibinfo {journal} {Appl. Phys. Lett.}}\ }%
  \textbf{\bibinfo {volume} {65}},\ \bibinfo {pages} {1085} (\bibinfo {month}
  {Aug.}\ \bibinfo {year} {1994}),\ ISSN \bibinfo {issn} {0003-6951,
  1077-3118}\BibitemShut{NoStop}%
\bibitem{vecchi2004}%
  \BibitemOpen
  \bibfield{author}{%
  \bibinfo {author} {\bibfnamefont{G.}~\bibnamefont{Vecchi}}, \bibinfo {author}
  {\bibfnamefont{J.}~\bibnamefont{Torres}}, \bibinfo {author}
  {\bibfnamefont{D.}~\bibnamefont{Coquillat}}, \bibinfo {author}
  {\bibfnamefont{M.~L.~V.}\ \bibnamefont{d’Yerville}},\ and\ \bibinfo
  {author} {\bibfnamefont{A.~M.}\ \bibnamefont{Malvezzi}},\ }%
  \bibfield{journal}{%
  \Doi{10.1063/1.1649800}{\bibinfo {journal} {Appl. Phys. Lett.}}\ }%
  \textbf{\bibinfo {volume} {84}},\ \bibinfo {pages} {1245} (\bibinfo {month}
  {Feb.}\ \bibinfo {year} {2004}),\ ISSN \bibinfo {issn} {0003-6951,
  1077-3118}\BibitemShut{NoStop}%
\bibitem{xiongpernice2011}%
  \BibitemOpen
  \bibfield{author}{%
  \bibinfo {author} {\bibfnamefont{C.}~\bibnamefont{Xiong}}, \bibinfo {author}
  {\bibfnamefont{W.}~\bibnamefont{Pernice}}, \bibinfo {author}
  {\bibfnamefont{K.~K.}\ \bibnamefont{Ryu}}, \bibinfo {author}
  {\bibfnamefont{C.}~\bibnamefont{Schuck}}, \bibinfo {author}
  {\bibfnamefont{K.~Y.}\ \bibnamefont{Fong}}, \bibinfo {author}
  {\bibfnamefont{T.}~\bibnamefont{Palacios}},\ and\ \bibinfo {author}
  {\bibfnamefont{H.~X.}\ \bibnamefont{Tang}},\ }%
  \bibfield{journal}{%
  \Doi{10.1364/OE.19.010462}{\bibinfo {journal} {Opt. Express}}\ }%
  \textbf{\bibinfo {volume} {19}},\ \bibinfo {pages} {10462–10470} (\bibinfo
  {month} {May}\ \bibinfo {year} {2011})\BibitemShut{NoStop}%
\bibitem{xiongpernice2012}%
  \BibitemOpen
  \bibfield{author}{%
  \bibinfo {author} {\bibfnamefont{C.}~\bibnamefont{Xiong}}, \bibinfo {author}
  {\bibfnamefont{W.~H.~P.}\ \bibnamefont{Pernice}}, \bibinfo {author}
  {\bibfnamefont{X.}~\bibnamefont{Sun}}, \bibinfo {author}
  {\bibfnamefont{C.}~\bibnamefont{Schuck}}, \bibinfo {author}
  {\bibfnamefont{K.~Y.}\ \bibnamefont{Fong}},\ and\ \bibinfo {author}
  {\bibfnamefont{H.~X.}\ \bibnamefont{Tang}},\ }%
  \bibfield{journal}{%
  \Doi{10.1088/1367-2630/14/9/095014}{\bibinfo {journal} {New J. Phys.}}\ }%
  \textbf{\bibinfo {volume} {14}},\ \bibinfo {pages} {095014} (\bibinfo {month}
  {Sep.}\ \bibinfo {year} {2012}),\ ISSN \bibinfo {issn} {1367-2630},\
  \url{http://iopscience.iop.org/1367-2630/14/9/095014}\BibitemShut{NoStop}%
\bibitem{stutzmann2006}%
  \BibitemOpen
  \bibfield{author}{%
  \bibinfo {author} {\bibfnamefont{M.}~\bibnamefont{Stutzmann}}, \bibinfo
  {author} {\bibfnamefont{J.~A.}\ \bibnamefont{Garrido}}, \bibinfo {author}
  {\bibfnamefont{M.}~\bibnamefont{Eickhoff}},\ and\ \bibinfo {author}
  {\bibfnamefont{M.~S.}\ \bibnamefont{Brandt}},\ }%
  \bibfield{journal}{%
  \Doi{10.1002/pssa.200622512}{\bibinfo {journal} {Phys. Status Solidi A}}\ }%
  \textbf{\bibinfo {volume} {203}},\ \bibinfo {pages} {3424} (\bibinfo {month}
  {Nov.}\ \bibinfo {year} {2006}),\ ISSN \bibinfo {issn}
  {1862-6319}\BibitemShut{NoStop}%
\bibitem{sergent2012}%
  \BibitemOpen
  \bibfield{author}{%
  \bibinfo {author} {\bibfnamefont{S.}~\bibnamefont{Sergent}}, \bibinfo
  {author} {\bibfnamefont{M.}~\bibnamefont{Arita}}, \bibinfo {author}
  {\bibfnamefont{S.}~\bibnamefont{Kako}}, \bibinfo {author}
  {\bibfnamefont{K.}~\bibnamefont{Tanabe}}, \bibinfo {author}
  {\bibfnamefont{S.}~\bibnamefont{Iwamoto}},\ and\ \bibinfo {author}
  {\bibfnamefont{Y.}~\bibnamefont{Arakawa}},\ }%
  \bibfield{journal}{%
  \Doi{10.1063/1.4751336}{\bibinfo {journal} {Appl. Phys. Lett.}}\ }%
  \textbf{\bibinfo {volume} {101}},\ \bibinfo {pages} {101106} (\bibinfo
  {month} {Sep.}\ \bibinfo {year} {2012}),\ ISSN \bibinfo {issn} {0003-6951,
  1077-3118},\
  \url{http://scitation.aip.org/content/aip/journal/apl/101/10/10.1063/1.47513%
36}\BibitemShut{NoStop}%
\bibitem{vicotrivino2012}%
  \BibitemOpen
  \bibfield{author}{%
  \bibinfo {author} {\bibfnamefont{N.}~\bibnamefont{{Vico Trivi\~{n}o}}},
  \bibinfo {author} {\bibfnamefont{G.}~\bibnamefont{Rossbach}}, \bibinfo
  {author} {\bibfnamefont{U.}~\bibnamefont{Dharanipathy}}, \bibinfo {author}
  {\bibfnamefont{J.}~\bibnamefont{Levrat}}, \bibinfo {author}
  {\bibfnamefont{A.}~\bibnamefont{Castiglia}}, \bibinfo {author}
  {\bibfnamefont{J.-F.}\ \bibnamefont{Carlin}}, \bibinfo {author}
  {\bibfnamefont{K.~A.}\ \bibnamefont{Atlasov}}, \bibinfo {author}
  {\bibfnamefont{R.}~\bibnamefont{Butt\'{e}}}, \bibinfo {author}
  {\bibfnamefont{R.}~\bibnamefont{Houdr\'{e}}},\ and\ \bibinfo {author}
  {\bibfnamefont{N.}~\bibnamefont{Grandjean}},\ }%
  \bibfield{journal}{%
  \Doi{doi:10.1063/1.3684630}{\bibinfo {journal} {Appl. Phys. Lett.}}\ }%
  \textbf{\bibinfo {volume} {100}},\ \bibinfo {pages} {071103} (\bibinfo
  {month} {Feb.}\ \bibinfo {year} {2012}),\ ISSN \bibinfo {issn} {00036951},\
  \url{http://apl.aip.org/resource/1/applab/v100/i7/p071103_s1}\BibitemShut{No%
Stop}%
\bibitem{samgiao2012}%
  \BibitemOpen
  \bibfield{author}{%
  \bibinfo {author} {\bibfnamefont{D.}~\bibnamefont{Sam-Giao}}, \bibinfo
  {author} {\bibfnamefont{D.}~\bibnamefont{N\'{e}el}}, \bibinfo {author}
  {\bibfnamefont{S.}~\bibnamefont{Sergent}}, \bibinfo {author}
  {\bibfnamefont{B.}~\bibnamefont{Gayral}}, \bibinfo {author}
  {\bibfnamefont{M.~J.}\ \bibnamefont{Rashid}}, \bibinfo {author}
  {\bibfnamefont{F.}~\bibnamefont{Semond}}, \bibinfo {author}
  {\bibfnamefont{J.~Y.}\ \bibnamefont{Duboz}}, \bibinfo {author}
  {\bibfnamefont{M.}~\bibnamefont{Mexis}}, \bibinfo {author}
  {\bibfnamefont{T.}~\bibnamefont{Guillet}}, \bibinfo {author}
  {\bibfnamefont{C.}~\bibnamefont{Brimont}}, \bibinfo {author}
  {\bibfnamefont{S.}~\bibnamefont{David}}, \bibinfo {author}
  {\bibfnamefont{X.}~\bibnamefont{Checoury}},\ and\ \bibinfo {author}
  {\bibfnamefont{P.}~\bibnamefont{Boucaud}},\ }%
  \bibfield{journal}{%
  \Doi{10.1063/1.4712590}{\bibinfo {journal} {Appl. Phys. Lett.}}\ }%
  \textbf{\bibinfo {volume} {100}},\ \bibinfo {pages} {191104} (\bibinfo
  {month} {May}\ \bibinfo {year} {2012}),\ ISSN \bibinfo {issn} {0003-6951,
  1077-3118},\
  \url{http://scitation.aip.org/content/aip/journal/apl/100/19/10.1063/1.47125%
90}\BibitemShut{NoStop}%
\bibitem{arita2012}%
  \BibitemOpen
  \bibfield{author}{%
  \bibinfo {author} {\bibfnamefont{M.}~\bibnamefont{Arita}}, \bibinfo {author}
  {\bibfnamefont{S.}~\bibnamefont{Kako}}, \bibinfo {author}
  {\bibfnamefont{S.}~\bibnamefont{Iwamoto}},\ and\ \bibinfo {author}
  {\bibfnamefont{Y.}~\bibnamefont{Arakawa}},\ }%
  \bibfield{journal}{%
  \Doi{10.1143/APEX.5.126502}{\bibinfo {journal} {Appl. Phys. Express}}\ }%
  \textbf{\bibinfo {volume} {5}},\ \bibinfo {pages} {126502} (\bibinfo {month}
  {Dec.}\ \bibinfo {year} {2012})\BibitemShut{NoStop}%
\bibitem{VicoTrivino2013}%
  \BibitemOpen
  \bibfield{author}{%
  \bibinfo {author} {\bibfnamefont{N.}~\bibnamefont{{Vico Trivi\~{n}o}}},
  \bibinfo {author} {\bibfnamefont{U.}~\bibnamefont{Dharanipathy}}, \bibinfo
  {author} {\bibfnamefont{J.-F.}\ \bibnamefont{Carlin}}, \bibinfo {author}
  {\bibfnamefont{Z.}~\bibnamefont{Diao}}, \bibinfo {author}
  {\bibfnamefont{R.}~\bibnamefont{Houdr\'{e}}},\ and\ \bibinfo {author}
  {\bibfnamefont{N.}~\bibnamefont{Grandjean}},\ }%
  \bibfield{journal}{%
  \Doi{10.1063/1.4793759}{\bibinfo {journal} {Applied Physics Letters}}\ }%
  \textbf{\bibinfo {volume} {102}},\ \bibinfo {pages} {081120} (\bibinfo {year}
  {2013}),\ ISSN \bibinfo {issn} {00036951},\
  \url{http://scitation.aip.org/content/aip/journal/apl/102/8/10.1063/1.479375%
9}\BibitemShut{NoStop}%
\bibitem{Roland2014}%
  \BibitemOpen
  \bibfield{author}{%
  \bibinfo {author} {\bibfnamefont{I.}~\bibnamefont{Roland}}, \bibinfo {author}
  {\bibfnamefont{Y.}~\bibnamefont{Zeng}}, \bibinfo {author}
  {\bibfnamefont{Z.}~\bibnamefont{Han}}, \bibinfo {author}
  {\bibfnamefont{X.}~\bibnamefont{Checoury}}, \bibinfo {author}
  {\bibfnamefont{C.}~\bibnamefont{Blin}}, \bibinfo {author}
  {\bibfnamefont{M.}~\bibnamefont{{El Kurdi}}}, \bibinfo {author}
  {\bibfnamefont{A.}~\bibnamefont{Ghrib}}, \bibinfo {author}
  {\bibfnamefont{S.}~\bibnamefont{Sauvage}}, \bibinfo {author}
  {\bibfnamefont{B.}~\bibnamefont{Gayral}}, \bibinfo {author}
  {\bibfnamefont{C.}~\bibnamefont{Brimont}}, \bibinfo {author}
  {\bibfnamefont{T.}~\bibnamefont{Guillet}}, \bibinfo {author}
  {\bibfnamefont{F.}~\bibnamefont{Semond}},\ and\ \bibinfo {author}
  {\bibfnamefont{P.}~\bibnamefont{Boucaud}},\ }%
  \bibfield{journal}{%
  \Doi{10.1063/1.4887065}{\bibinfo {journal} {Appl. Phys. Lett.}}\ }%
  \textbf{\bibinfo {volume} {105}},\ \bibinfo {pages} {011104} (\bibinfo
  {month} {Jul.}\ \bibinfo {year} {2014})\BibitemShut{NoStop}%
\bibitem{pernice2012}%
  \BibitemOpen
  \bibfield{author}{%
  \bibinfo {author} {\bibfnamefont{W.~H.~P.}\ \bibnamefont{Pernice}}, \bibinfo
  {author} {\bibfnamefont{C.}~\bibnamefont{Xiong}}, \bibinfo {author}
  {\bibfnamefont{C.}~\bibnamefont{Schuck}},\ and\ \bibinfo {author}
  {\bibfnamefont{H.~X.}\ \bibnamefont{Tang}},\ }%
  \bibfield{journal}{%
  \Doi{doi:10.1063/1.3690888}{\bibinfo {journal} {Appl. Phys. Lett.}}\ }%
  \textbf{\bibinfo {volume} {100}},\ \bibinfo {pages} {091105} (\bibinfo
  {month} {Feb.}\ \bibinfo {year} {2012}),\ ISSN \bibinfo {issn} {00036951},\
  \url{http://apl.aip.org/resource/1/applab/v100/i9/p091105_s1}\BibitemShut{No%
Stop}%
\bibitem{Portalupi2011}%
  \BibitemOpen
  \bibfield{author}{%
  \bibinfo {author} {\bibfnamefont{S.~L.}\ \bibnamefont{Portalupi}}, \bibinfo
  {author} {\bibfnamefont{M.}~\bibnamefont{Galli}}, \bibinfo {author}
  {\bibfnamefont{M.}~\bibnamefont{Belotti}}, \bibinfo {author}
  {\bibfnamefont{L.~C.}\ \bibnamefont{Andreani}}, \bibinfo {author}
  {\bibfnamefont{T.~F.}\ \bibnamefont{Krauss}},\ and\ \bibinfo {author}
  {\bibfnamefont{L.}~\bibnamefont{{O'Faolain}}},\ }%
  \bibfield{journal}{%
  \Doi{10.1103/PhysRevB.84.045423}{\bibinfo {journal} {Phys. Rev. B}}\ }%
  \textbf{\bibinfo {volume} {84}},\ \bibinfo {pages} {045423} (\bibinfo {month}
  {Jul.}\ \bibinfo {year} {2011}),\
  \url{http://link.aps.org/doi/10.1103/PhysRevB.84.045423}\BibitemShut{NoStop}%
\bibitem{Minkov2013}%
  \BibitemOpen
  \bibfield{author}{%
  \bibinfo {author} {\bibfnamefont{M.}~\bibnamefont{Minkov}}, \bibinfo {author}
  {\bibfnamefont{U.~P.}\ \bibnamefont{Dharanipathy}}, \bibinfo {author}
  {\bibfnamefont{R.}~\bibnamefont{Houdr\'{e}}},\ and\ \bibinfo {author}
  {\bibfnamefont{V.}~\bibnamefont{Savona}},\ }%
  \bibfield{journal}{%
  \bibinfo {journal} {Opt. Express}\ }%
  \textbf{\bibinfo {volume} {21}},\ \bibinfo {pages} {28233} (\bibinfo {month}
  {Nov}\ \bibinfo {year} {2013})\BibitemShut{NoStop}%
\bibitem{Minkov2014}%
  \BibitemOpen
  \bibfield{author}{%
  \bibinfo {author} {\bibfnamefont{M.}~\bibnamefont{Minkov}}\ and\ \bibinfo
  {author} {\bibfnamefont{V.}~\bibnamefont{Savona}},\ }%
  \bibfield{journal}{%
  \Doi{10.1038/srep05124}{\bibinfo {journal} {Scientific reports}}\ }%
  \textbf{\bibinfo {volume} {4}},\ \bibinfo {pages} {5124} (\bibinfo {month}
  {Jan.}\ \bibinfo {year} {2014}),\ ISSN \bibinfo {issn} {2045-2322},\
  \url{http://www.pubmedcentral.nih.gov/articlerender.fcgi?artid=4038819\&tool%
=pmcentrez\&rendertype=abstract}\BibitemShut{NoStop}%
\bibitem{vincent2003}%
  \BibitemOpen
  \bibfield{author}{%
  \bibinfo {author} {\bibfnamefont{N.}~\bibnamefont{Antoine-Vincent}}, \bibinfo
  {author} {\bibfnamefont{F.}~\bibnamefont{Natali}}, \bibinfo {author}
  {\bibfnamefont{M.}~\bibnamefont{Mihailovic}}, \bibinfo {author}
  {\bibfnamefont{A.}~\bibnamefont{Vasson}}, \bibinfo {author}
  {\bibfnamefont{J.}~\bibnamefont{Leymarie}}, \bibinfo {author}
  {\bibfnamefont{P.}~\bibnamefont{Disseix}}, \bibinfo {author}
  {\bibfnamefont{D.}~\bibnamefont{Byrne}}, \bibinfo {author}
  {\bibfnamefont{F.}~\bibnamefont{Semond}},\ and\ \bibinfo {author}
  {\bibfnamefont{J.}~\bibnamefont{Massies}},\ }%
  \bibfield{journal}{%
  \Doi{10.1063/1.1563293}{\bibinfo {journal} {J. Appl. Phys.}}\ }%
  \textbf{\bibinfo {volume} {93}},\ \bibinfo {pages} {5222} (\bibinfo {month}
  {May}\ \bibinfo {year} {2003}),\ ISSN \bibinfo {issn} {0021-8979,
  1089-7550}\BibitemShut{NoStop}%
\bibitem{Andreani2006}%
  \BibitemOpen
  \bibfield{author}{%
  \bibinfo {author} {\bibfnamefont{L.~C.}\ \bibnamefont{Andreani}}\ and\
  \bibinfo {author} {\bibfnamefont{D.}~\bibnamefont{Gerace}},\ }%
  \bibfield{journal}{%
  \Doi{10.1103/PhysRevB.73.235114}{\bibinfo {journal} {Phys. Rev. B}}\ }%
  \textbf{\bibinfo {volume} {73}},\ \bibinfo {pages} {235114} (\bibinfo {month}
  {Jun.}\ \bibinfo {year} {2006}),\
  \url{http://link.aps.org/doi/10.1103/PhysRevB.73.235114}\BibitemShut{NoStop}%
\bibitem{matlab}%
  \BibitemOpen
  \enquote{\bibinfo {title} {{MATLAB} and global optimization toolbox release
  2012b},}\  (\bibinfo {year} {2012})\BibitemShut{NoStop}%
\bibitem{Lai2014}%
  \BibitemOpen
  \bibfield{author}{%
  \bibinfo {author} {\bibfnamefont{Y.}~\bibnamefont{Lai}}, \bibinfo {author}
  {\bibfnamefont{S.}~\bibnamefont{Pirotta}}, \bibinfo {author}
  {\bibfnamefont{G.}~\bibnamefont{Urbinati}}, \bibinfo {author}
  {\bibfnamefont{D.}~\bibnamefont{Gerace}}, \bibinfo {author}
  {\bibfnamefont{M.}~\bibnamefont{Minkov}}, \bibinfo {author}
  {\bibfnamefont{V.}~\bibnamefont{Savona}}, \bibinfo {author}
  {\bibfnamefont{A.}~\bibnamefont{Badolato}},\ and\ \bibinfo {author}
  {\bibfnamefont{M.}~\bibnamefont{Galli}},\ }%
  \bibfield{journal}{%
  \Doi{10.1063/1.4882860}{\bibinfo {journal} {Applied Physics Letters}}\ }%
  \textbf{\bibinfo {volume} {104}},\ \bibinfo {pages} {241101} (\bibinfo
  {month} {Jun.}\ \bibinfo {year} {2014}),\ ISSN \bibinfo {issn} {0003-6951},\
  \url{http://scitation.aip.org/content/aip/journal/apl/104/24/10.1063/1.48828%
60}\BibitemShut{NoStop}%
\bibitem{Dharanipathy2014}%
  \BibitemOpen
  \bibfield{author}{%
  \bibinfo {author} {\bibfnamefont{U.~P.}\ \bibnamefont{Dharanipathy}},
  \bibinfo {author} {\bibfnamefont{M.}~\bibnamefont{Minkov}}, \bibinfo {author}
  {\bibfnamefont{M.}~\bibnamefont{Tonin}}, \bibinfo {author}
  {\bibfnamefont{V.}~\bibnamefont{Savona}},\ and\ \bibinfo {author}
  {\bibfnamefont{R.}~\bibnamefont{Houdr\'{e}}},\ }%
  \bibfield{journal}{%
  \Doi{10.1063/1.4894441}{\bibinfo {journal} {Appl. Phys. Lett.}}\ }%
  \textbf{\bibinfo {volume} {105}},\ \bibinfo {pages} {101101} (\bibinfo
  {month} {Sep.}\ \bibinfo {year} {2014}),\ ISSN \bibinfo {issn} {0003-6951},\
  \url{http://scitation.aip.org/content/aip/journal/apl/105/10/10.1063/1.48944%
41}\BibitemShut{NoStop}%
\bibitem{lumerical}%
  \BibitemOpen
  \enquote{\bibinfo {title} {Lumerical solutions, inc.}.}\
  \url{http://www.lumerical.com/tcad-products/mode/}\BibitemShut{NoStop}%
\bibitem{Arita2007}%
  \BibitemOpen
  \bibfield{author}{%
  \bibinfo {author} {\bibfnamefont{M.}~\bibnamefont{Arita}}, \bibinfo {author}
  {\bibfnamefont{S.}~\bibnamefont{Ishida}}, \bibinfo {author}
  {\bibfnamefont{S.}~\bibnamefont{Kako}}, \bibinfo {author}
  {\bibfnamefont{S.}~\bibnamefont{Iwamoto}},\ and\ \bibinfo {author}
  {\bibfnamefont{Y.}~\bibnamefont{Arakawa}},\ }%
  \bibfield{journal}{%
  \Doi{10.1063/1.2757596}{\bibinfo {journal} {Appl. Phys. Lett.}}\ }%
  \textbf{\bibinfo {volume} {91}},\ \bibinfo {pages} {051106} (\bibinfo {year}
  {2007}),\ ISSN \bibinfo {issn} {00036951},\
  \url{http://scitation.aip.org/content/aip/journal/apl/91/5/10.1063/1.2757596%
}\BibitemShut{NoStop}%
\bibitem{Minkov2012}%
  \BibitemOpen
  \bibfield{author}{%
  \bibinfo {author} {\bibfnamefont{M.}~\bibnamefont{Minkov}}\ and\ \bibinfo
  {author} {\bibfnamefont{V.}~\bibnamefont{Savona}},\ }%
  \bibfield{journal}{%
  \Doi{10.1364/OL.37.003108}{\bibinfo {journal} {Opt. Lett.}}\ }%
  \textbf{\bibinfo {volume} {37}},\ \bibinfo {pages} {3108} (\bibinfo {month}
  {Aug}\ \bibinfo {year} {2012}),\
  \url{http://ol.osa.org/abstract.cfm?URI=ol-37-15-3108}\BibitemShut{NoStop}%
\end{thebibliography}

%

\end{document}